\documentclass[twocolumn,showpacs,preprintnumbers,amsmath,amssymb]{revtex4}

\usepackage{graphicx}
\usepackage{amssymb,graphics,pstricks,pst-node,pst-char,pst-plot,pst-3d,amsmath}
\newpsobject{showgrid}{psgrid}{subgriddiv=5}

\usepackage{dcolumn}
\usepackage{bm}

\begin{document}

\title{Sensitivity to initial conditions in self-organized critical systems}

\author{Matthew Stapleton}
\email{matthew.stapleton@imperial.ac.uk}
\author{Martin Dingler}
\author{Kim Christensen}
\email{k.christensen@imperial.ac.uk}
\homepage{http://www.imperial.ac.uk/research/cmth/people/k.christensen/}
\affiliation{Blackett Laboratory, Imperial College London,
Prince Consort Road, London SW7 2BW, United Kingdom}

\date{\today}

\begin{abstract}
We discuss sensitivity to initial conditions in a model for avalanches in granular media
displaying self-organized criticality.
We show that damage, due to a small perturbation in initial conditions, does not spread.
The damage persists in a statistically time-invariant and scale-free form.
We argue that the origin of this behavior is the Abelian nature of the model,
which generalizes our results to all models with Abelian properties, including the BTW model
and the Manna model.
An ensemble average of the damage leads to seemingly time dependent damage spreading.
Scaling arguments show that this numerical result 
is due to the time lag before avalanches reach the initial perturbation.
\end{abstract}

\maketitle

There is evidence that the dynamics of a pile of rice may display self-organized criticality (SOC) \cite{BTW87}.
In careful experiments where elongated rice grains were slowly dropped
between two glass plates, Frette {\it et al.} found scale-free behavior
in a rice pile \cite{Nature}.
In a slowly driven pile, the angle of repose evolves to a stationary state
and the behavior of the system is dominated by a scale-free avalanche size
density and punctuated equilibrium.
This punctuation causes SOC systems to be highly non-linear, a single
grain added at one end can result in an avalanche propagating through the
entire system.
However, SOC models typically have stick-slip dynamics \cite{HJensen} and
it has yet to be established whether they allow the non-linearity to
manifest itself in the form of sensitivity to initial conditions, as it
does in chaotic systems.

We have studied damage spreading in a simple one-dimensional granular model known as the Oslo
model, which exhibits SOC \cite{Oslo}.
It describes a number of slowly driven granular systems and belongs to the same universality
class as a model for interface depinning in a random medium and the Burridge-Knopoff
train model for earthquakes \cite{MPacz,BK,Train}.
The Oslo model has largely resisted efforts for an analytic solution,
the few exceptions have been the exact enumeration of the number of recurrent configurations 
\cite{Chua}, the mapping of the model to the quenched Edwards-Wilkinson
equation in the continuous limit \cite{MPacz,qEW}, and the transition matrix results 
\cite{Corral} and operator algebra recently developed for the Oslo model \cite{Dhar2}.
In this letter, we add to its analytical description by illustrating its Abelian
properties \cite{Dhar}.

We find that damage due to a small perturbation does not spread in the Oslo
model as the damage is unable to evolve.
It is possible to represent the perturbation in terms of commutative operators which 
leads to a statistically time-invariant and scale-free damage.
This phenomenon is in contrast to what is seen in chaotic and equilibrium systems.
We also address the results obtained by a previous study on an ensemble average
of the damage, which seems to contradict our findings \cite{Gleiser}.
In fact, we show that these results are consistent with ours and that the observed behavior
may be derived using simple scaling arguments.

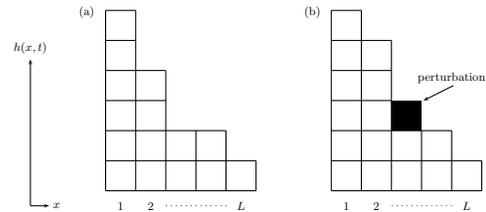
\begin{figure}[h]
\begin{pspicture}(-1.5,0)(15,2.25)
{\scalebox{0.5}{
\psline(1.0,4.8)(1.8,4.8) \psline(1.0,0.0)(1.0,4.8)
\psline(1.0,4.0)(1.8,4.0) \psline(1.8,0.0)(1.8,4.8)
\psline(1.0,3.2)(2.6,3.2) \psline(2.6,0.0)(2.6,3.2)
\psline(1.0,2.4)(2.6,2.4) \psline(3.4,0.0)(3.4,1.6)
\psline(1.0,1.6)(4.2,1.6) \psline(4.2,0.0)(4.2,1.6)
\psline(1.0,0.8)(5.0,0.8) \psline(5.0,0.0)(5.0,0.8)
\psline(1.0,0.0)(5.0,0.0)

\psline(7.0,0.0)(7.0,4.8) \psline(7.0,4.8)(7.8,4.8)
\psline(7.0,4.0)(8.6,4.0) \psline(7.8,0.0)(7.8,4.8)
\psline(7.0,3.2)(8.6,3.2) \psline(8.6,0.0)(8.6,4.0)
\psline(7.0,2.4)(8.6,2.4) \psline(9.4,0.0)(9.4,1.6)
\psline(7.0,1.6)(10.2,1.6) \psline(10.2,0.0)(10.2,1.6)
\psline(7.0,0.8)(11.0,0.8) \psline(11.0,0.0)(11.0,0.8)
\psline(7.0,0.0)(11.0,0.0)
\psframe[fillstyle=solid,fillcolor=black](8.6,1.6)(9.4,2.4)

\rput(10.2,3.0){perturbation}
\psline{->}(10.25,2.8)(9.45,2.4)
\rput(0.5,4.75){(a)}
\rput(6.5,4.75){(b)}
\rput(1.4,-0.4){$1$} \rput(2.2,-0.4){$2$} \psline[linestyle=dotted](2.6,-0.4)(4.2,-0.4) \rput(4.6,-0.4){$L$}
\rput(7.4,-0.4){$1$} \rput(8.2,-0.4){$2$} \psline[linestyle=dotted](8.6,-0.4)(10.2,-0.4) \rput(10.6,-0.4){$L$}
\psline{->}(-1.0,-0.4)(-0.5,-0.4)  \rput(-0.3,-0.4){$x$}
\psline{->}(-1,-0.4)(-1,3.5) \rput(-1,3.8){$h(x,t)$}
}}
\end{pspicture}
\caption{
(a) The Oslo model of a one-dimensional granular pile. Grains are added at the
site $x=1$ next to the vertical wall by letting $h(1,t+1) = h(1,t) + 1$.
A grain at site $x$ topples to site $x+1$ if the local slope $z(x,t) = h(x,t) - h(x+1,t)$
exceeds its critical slope $z_c(x)$.
When $z(L,t) > z_c(L)$, a grain leaves the system at the open boundary, $h(L,t) \rightarrow h(L,t) - 1$. 
(b) The  solid grain is the additional grain in the copy at time $t_0$.}
\label{F:1dlattice}
\end{figure}
{\em The model:}
The Oslo model is defined on a one dimensional discrete lattice with
$L$ sites at positions $x = 1,2,\ldots L$, see Fig.~\ref{F:1dlattice}.
On the left, the boundary is a vertical wall, and on the right, the boundary is open.
The height of grains at site $x$ and time $t$ is denoted $h(x,t)$, the local slope
is defined as $z(x,t) = h(x,t) - h(x+1,t)$.
Each site has a critical slope, $z_c(x)$, which takes the values $1$ or $2$ with equal probability.
At each time step a single grain is added to the site at $x = 1$.
If the local slope at any site, $x$, exceeds its critical slope, $z(x,t) > z_c(x)$, an
avalanche is initiated.
The site will relax and a grain will topple from site $x$ to site $x+1$, i.e. $h(x,t) \rightarrow h(x,t) - 1$
and $h(x+1,t) \rightarrow h(x+1,t) + 1$.
Each time a site relaxes its critical slope is redetermined, chosen randomly from the values $1$ or $2$.
This toppling may cause sites $x\pm 1$ to exceed their critical slopes, in which case these sites
relax in turn.
The avalanche will continue until the system reaches a  stable configuration, when $z(x,t) \leq z_c(x)$ for
all $x$.

In order to study damage spreading in the model we consider a system which has been evolved
to the critical state and make an exact copy at some time, $t_0$.
We define $h^o(x,t)$ and $h^c(x,t)$ as the heights of the original and copy,
respectively, and $z^o_c(x)$ and $z^c_c(x)$ as the critical slopes of the
original and copy, such that
\begin {equation}
\left.
\begin{matrix}
h^c(x,t_0) & = & h^o(x,t_0) \\
z^c_c(x) & = & z^o_c(x)
\end{matrix} \right. \qquad
1 \leq x \leq L.
\end {equation}
We then perturb the copy by adding a single grain to a site, $i$,
such that $h^c(i,t_0) = h^o(i,t_0) + 1$, see Fig.~\ref{F:1dlattice}(b).
This extra grain is not allowed to topple until it is toppled upon by an avalanche from above.
The two systems are then evolved using exactly the same sequence of random thresholds
$\{ z_c(x) \}$ for corresponding sites in the original and copy.
This ensures that the damage observed is purely due to the extra grain in the copy.
Further justification for this is to
consider the mapping of the model to interface depinning as it is clear that the medium
the interface moves through does not change as a result of the perturbation.

We measure the damage, defined as
\begin {equation}
H(t,L) = \sum^L_{x=1} |h^o(x,t) - h^c(x,t)|.
\end {equation}
\begin{figure}[h]
\begin{pspicture}(1.25,1.85)(11,5.1)
\rput(5.5,3.5){\scalebox{0.35}{\includegraphics{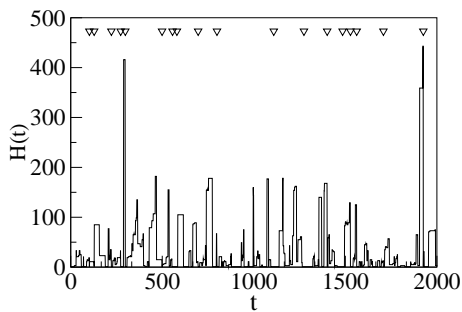}}}
\end{pspicture}
\caption{Damage as a function of time for a single run, $L = 128$, where the perturbed
site is $i = 16$ and we have taken $t_0 = 0$.
Notice that there is no temporal evolution in the data, the damage is continually
fluctuating between low and high values and frequently returns to a $H = 1$ configuration.
The triangles indicate where the $H=1$ configuration corresponds to a repaired configuration, where corresponding
sites in the original and copy have relaxed an equal number of times.}
\label{F:linear}
\end{figure}

Figure~\ref{F:linear} is a plot of damage versus time for a typical simulation.
There appears to be no sense of temporal evolution in the data, the damage
is continually fluctuating between high and low values, often returning to
$H = 1$ where the original and copy only differ by a single grain.
The triangles in Fig.~\ref{F:linear} indicate which of these $H = 1$ configurations are repaired configurations where the extra
grain in the copy is at the site which was originally perturbed
and corresponding sites in the original and copy have relaxed exactly the same
number of times.
In repaired configurations, the difference between the
two systems is {\it exactly equivalent} to the initial perturbation at $t_0$.
The occurrence of a repaired configuration corresponds to `resetting the clock',
meaning that the damage may not evolve in a single pair of systems, it is statistically
time invariant.

This behavior emerges because the Oslo model has an Abelian nature,
where the commutative operation is adding one unit of {\em slope} to a
site and allowing the system to relax.
If we add one unit of slope at site $x$ and then at site $y$, it is equivalent
to adding at site $y$ and then $x$.
The Abelian nature is alluded to by the fact that these simulations require
a careful record of the sequences of random critical slopes.
In this way the sequences of critical slopes are no longer random noise,
but an intrinsic property of the system where the values, although generated randomly,
are treated as given {\it a priori}.
Note that the model is not Abelian in the strict sense of an Abelian group as there is no inverse operation.

We introduce the notation $C(t)$ to represent the stable configuration of a system at time $t$.
The relevant operators are the perturbation operator $P_i$, which simply adds one unit of slope to
site $i$ and $\hat P_i$, which adds one unit of slope to site $i$ {\it and} allows the system to relax
if necessary.
Thus, $\hat P_i$ represents a mapping within the configuration space of $C(t)$.
The evolution of $C(t)$ can be expressed by an evolution operator $\hat T \equiv \hat P_1$, that is
\begin {equation}
C(t+1) = \hat TC(t).
\end {equation}
The operators $\hat P_i$ and $\hat T$ obey the commutation relation
\begin {equation}
\hat P_i \hat T C(t) = \hat T \hat P_i C(t), \label{eq:comm}
\end {equation}
which we prove elsewhere \cite{next_one}.
Hence, adding slope to a system and then evolving it leads to the same configuration as
evolving it and then adding slope, as we may always move the $\hat P_i$ operator
on the right hand side of Eq.~(\ref{eq:comm}) to the left of the $\hat T$ operator.

If we begin with a system at a point $C(0)$ in configuration space, then the repeated
operation of $\hat T$ defines the chain of configurations which $C(t)$ will pass through during
its evolution.
Consider a succession of points $\tilde C(t)$, which is related to $C(t)$ by the relation
$\tilde C(0) \equiv \hat P_i C(0)$.  The points $C(t)$ and
$\tilde C(t)$ will be connected through the operator $\hat P_i$ at every step of
the evolution of the two systems, i.e. $\tilde C(t) \equiv \hat P_i C(t)$, for all $t$.
 
The perturbation we studied in the simulations was the addition of a single grain and
not slope.
Adding a single grain to site $i$ will change the slopes such that 
$z^c(i,t_0) = z^o(i,t_0) + 1$ and $z^c(i-1,t_0) = z^o(i-1,t_0) - 1$.
To analyze this situation, we start with a master system $C^m(0)$
and derive from this two other systems $C^o(0)$ and $C^c(0)$ through the relations
\begin {equation}
\begin {array}{rcl}
C^o(0) &=& P_{i-1} C^m(0) \\
C^c(0) &=& P_{i} C^m(0).
\end {array} \label{eq:relation}
\end {equation}
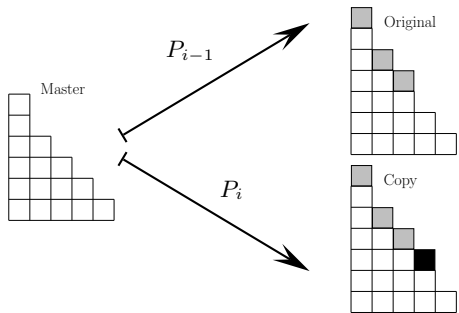
\begin{figure}[h]
\begin{pspicture}(1,0.75)(15,4)
{\scalebox{0.35}{
\rput(5,4.5){
\psline(0.0,4.8)(0.8,4.8) \psline(0.0,0.0)(0.0,4.8)
\psline(0.0,4.0)(0.8,4.0) \psline(0.8,0.0)(0.8,4.8)
\psline(0.0,3.2)(1.6,3.2) \psline(1.6,0.0)(1.6,3.2)
\psline(0.0,2.4)(2.4,2.4) \psline(2.4,0.0)(2.4,2.4)
\psline(0.0,1.6)(3.2,1.6) \psline(3.2,0.0)(3.2,1.6)
\psline(0.0,0.8)(4.0,0.8) \psline(4.0,0.0)(4.0,0.8)
\psline(0.0,0.0)(4.0,0.0)
\rput(2.25,5.0){{\LARGE{Master\phantom{ei}}}}
}

\rput(18,7){
\psline(0.0,4.8)(0.8,4.8) \psline(0.0,0.0)(0.0,4.8)
\psline(0.0,4.0)(0.8,4.0) \psline(0.8,0.0)(0.8,4.8)
\psline(0.0,3.2)(1.6,3.2) \psline(1.6,0.0)(1.6,3.2)
\psline(0.0,2.4)(2.4,2.4) \psline(2.4,0.0)(2.4,2.4)
\psline(0.0,1.6)(3.2,1.6) \psline(3.2,0.0)(3.2,1.6)
\psline(0.0,0.8)(4.0,0.8) \psline(4.0,0.0)(4.0,0.8)
\psline(0.0,0.0)(4.0,0.0)
\rput(2.25,5.0){{\LARGE{Original}}}
\psframe[fillstyle=solid,fillcolor=lightgray](0.0,4.8)(0.8,5.6)
\psframe[fillstyle=solid,fillcolor=lightgray](0.8,3.2)(1.6,4.0)
\psframe[fillstyle=solid,fillcolor=lightgray](1.6,2.4)(2.4,3.2)}

\rput(18,1){
\psline(0.0,4.8)(0.8,4.8) \psline(0.0,0.0)(0.0,4.8)
\psline(0.0,4.0)(0.8,4.0) \psline(0.8,0.0)(0.8,4.8)
\psline(0.0,3.2)(1.6,3.2) \psline(1.6,0.0)(1.6,3.2)
\psline(0.0,2.4)(2.4,2.4) \psline(2.4,0.0)(2.4,2.4)
\psline(0.0,1.6)(3.2,1.6) \psline(3.2,0.0)(3.2,1.6)
\psline(0.0,0.8)(4.0,0.8) \psline(4.0,0.0)(4.0,0.8)
\psline(0.0,0.0)(4.0,0.0)
\rput(2.25,5.0){{\LARGE{Copy\phantom{eee}}}}
\psframe[fillstyle=solid,fillcolor=lightgray](0.0,4.8)(0.8,5.6)
\psframe[fillstyle=solid,fillcolor=lightgray](0.8,3.2)(1.6,4.0)
\psframe[fillstyle=solid,fillcolor=lightgray](1.6,2.4)(2.4,3.2)
\psframe[fillstyle=solid,fillcolor=black](2.4,1.6)(3.2,2.4)}
}}
\pcline[arrowsize=0.25,arrowlength=2.0,arrowinset=0.4]{|->}(3.25,2.7)(5.75,4.2) \naput{$P_{i-1}$}
\pcline[arrowsize=0.25,arrowlength=2.0,arrowinset=0.4]{|->}(3.25,2.4)(5.75,0.9) \naput{$P_i$}
\end{pspicture}
\caption{
The relationship between the master, $C^m(0)$, original, $C^{o}(0)$, and copy, $C^c(0)$
configurations at time $t_0 = 0$, see Eq.~(\ref{eq:relation}). 
The shaded grains are those added to the master configuration to produce configuration
of the original system.  The solid grain is the additional grain added to the 
configuration of the copy at time $t_0$.}
\label{F:network}
\end{figure}
It is straightforward to verify that the configurations $C^o(0)$ and $C^c(0)$ differ by one grain at
site $i$, thus reproducing the original and perturbed systems of our simulations
with the perturbative grain placed at site $i$, see Fig.~\ref{F:network}.
Each of these points start a chain of stable configurations, linked by the operator $\hat T$.
After an avalanche has reached site $i$, each configuration in the master chain links to those belonging to
the original and copy by the operators $\hat P_{i-1}$ and $\hat P_i$, respectively, due to Eq.~(\ref{eq:comm}).
Note that the operator $\hat P_i$ does not have a unique inverse so there is no direct path between the
original and copy.

The important result from this is that damage cannot {\it spread}.
Spreading implies that the small amount of damage at $t_0$ leads to a little
more damage at $t_0 + 1$, and so on until the damage saturates the system, which is the case for a chaotic system.
The Abelian nature means that the value of the damage at any time is independent
of when the perturbation took place.
Hence, spreading is not possible.
However, the damage does not remain constant or decay, as in an equilibrium system, 
because the damage is exactly that due to the avalanches which would be caused by adding
slope to different sites in the master system at that time.
This leads to a damage size density that is related to the avalanche size density, 
as easily recognized by considering the special case of perturbing site $i = 1$.
As the avalanche size density is scale free, we find that the damage size density
is scale free also. In an infinite system the damage may become arbitrarily large,
yet it will frequently return to an $H = 1$ configuration! 
In other words such a system may be considered as lying on the edge of chaos \cite{PBak, Kauffman}.

The damage in a single pair of systems is statistically time invariant,
yet a previous study has found that the ensemble averaged data is not \cite{Gleiser}.
The ensemble average of damage over $N$ runs, gives the
average damage as a function of time
\begin {equation}
\langle H \rangle (t,L) = \frac{1}{N} \sum_{j=1}^N H_j(t,L,i_j),
\end {equation}
where $H_j(t,L,i_j)$ is the damage from a single run and
the variable $i_j$ is the site of perturbation for the $jth$ run.
In Ref.~\cite{Gleiser} it was found that for $i_j = L/2 \ \forall j$
\begin {equation}
\langle H \rangle (t,L) = t^z \ G\left(\frac{t}{L^{\beta}}\right),
\end {equation}
where $z$ and $\beta$ are exponents to be determined and $G(x)$ is constant for $x \ll 1$
and proportional to $x^{-z}$ for $x \gg 1$.
The apparent time dependence arises from the fact that
the perturbed site is not allowed to relax until it is relaxed upon from above.
This forces all the systems into a repaired configuration at the start of the simulation and
the average damage increases over time as more systems have avalanches which reach 
the perturbed site.

In the case where the perturbative grain was placed on
a random site for each system in the ensemble, scaling arguments may be used to derive the 
temporal evolution of $\langle H \rangle (t,L)$.
The derived equation agrees well with the simulation data.
First, we calculate how long avalanches take to reach the perturbed site,
which we denote as site $i$.
The linear avalanche size, $l$, is known to be related to the avalanche size, $s$, by
$s \propto l^D$, where $D$ is the avalanche fractal dimension, $D \approx 2.25$
\cite{MPacz}.
Assuming $s = l^D$, the probability $P_l(l,L)dl$ of having an avalanche with linear length in the range  $l \rightarrow l + dl$, 
obeys $P_l(l,L)\ dl = P_s(s,L)\ ds$,
where $P_s(s,L)ds$ is the probability of having an avalanche of size $s$ in the range $s \rightarrow s + ds$,
given by the scaling ansatz
\begin {equation}
P_s(s,L)\ ds = s^{-\tau} \mathcal{G}_s\left(\frac{s}{L^D}\right)\ ds,
\end {equation}
where $\mathcal{G}_s(x)$ is constant for $x \ll 1$ and decays rapidly for $x \gg 1$, and
$\tau$ is the avalanche exponent, $\tau \approx 1.55$ \cite{MPacz}.
We find
\begin {equation}
P_l(l,L)\ dl = l^{-D} \mathcal{G}_l\left(\frac{l}{L}\right)\ dl,
\end {equation}
where $\mathcal{G}_l(x)$ is constant for $x \ll 1$ and decays rapidly for $x \rightarrow 1$.
Note that we have used the scaling relation $\tau = 2 - 1/D$ which is derived from the fact that
$\langle s \rangle = L$\footnote{One might conjecture that the values for the critical exponents are
$\tau = \frac{14}{9}$, $D = \frac{9}{4}$ and $\chi = \frac{5}{4}$.}.

This result allows us to calculate the probability of having an avalanche of linear size
larger than some distance $X$, which we denote $\phi(X)$.
\begin {equation}
\phi(X) = \int_X^\infty P_l(l,L)\ dl \propto X^{1-D}.
\end {equation}
Hence, we may expect to have an avalanche of size $l > X$ within the timescale
\begin {equation}
t = \frac{1}{\phi(X)} = X^{D-1} = X^{\chi}, \label{eq:t_X}
\end {equation}
where we have used the scaling relation $\chi = D - 1$, relating the roughness exponent $\chi$
to the fractal dimension $D$ \cite{MPacz}.
We use Eq.~(\ref{eq:t_X}) to obtain an ansatz for $\langle X\rangle (t,L)$, the average linear distance reached by
the avalanches in a time t,
\begin {equation}
\langle X\rangle (t,L) = t^\frac{1}{\chi} g\left(\frac{t}{t_L}\right),
\end {equation}
where $t_L$ is the timescale after which the avalanches can be expected to have spanned the
entire system and $g(x)$ is constant for $x \ll 1$ and proportional to $x^{-\frac{1}{\chi}}$
for $x \gg 1$ which ensures that $\langle X\rangle (t,L) \leq L$ for all $t$.
By inserting $X =L$ into Eq.~(\ref{eq:t_X}) we see that $t_L \propto L^{\chi}$,
which leads to $\langle X \rangle (\infty ,L) = L$ as expected.

The average damage as a function of time and system size, $\langle H \rangle(t,L)$,
may therefore be expressed as
\begin {eqnarray}
\langle H \rangle(t,L) &=& L^{\alpha} F_{H \neq 1}(t,L) + \bigl(1 - F_{H \neq 1}(t,L)\bigr) \nonumber \\
& \approx & L^{\alpha} F_{H \neq 1}(t,L) \quad \hbox{for $L \gg 1$},
\end {eqnarray}
where $F_{H \neq 1}(t,L)$ is the fraction of systems in an $H \neq 1$ damaged configuration at
time $t$, $L^{\alpha}$ is the mean amount of damage of a system in the damaged configuration and the last approximation is true for large L.
If the positions of the perturbed sites are distributed uniformly among the systems in 
the ensemble, we expect
\begin {equation}
\langle H \rangle (t,L) = L^{\alpha} \frac{\langle X\rangle (t,L)}{L}
= t^\frac{\alpha}{\chi} G\left(\frac{t}{L^{\chi}}\right),\label{eq:power_random}
\end {equation}
where $G(x) = x^\frac{1 - \alpha}{\chi} g(x)$ and we have taken $t_0 = 0$.
Scaling arguments, taking care to use the scaling relation $\tau = 2 - 2/D$ for 
a bulk driven system \cite{Kim}, yields $\alpha = 1$. This is consistent with our
measurement of  $\alpha \approx 1$ and so, putting $\chi = 5/4$ and $\alpha = 1$
into Eq.~(\ref{eq:power_random}), we have
\begin {equation}
\langle H \rangle(t,L) = t^{0.80}G\Bigl(\frac{t}{L^{1.25}}\Bigr). \label{eq:no7}
\end {equation}
\begin{figure}[h]
\begin{pspicture}(1.25,1.75)(11,5)
\rput(5.5,3.5){\scalebox{0.35}{\includegraphics{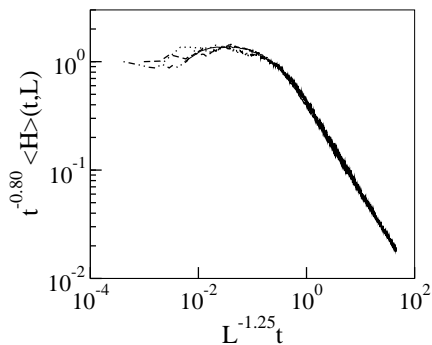}}}
\end{pspicture}
\caption{By plotting $t^{-0.80}\langle H \rangle (t,L)$ versus the rescaled time $x = L^{-1.25}t$ the data for the ensemble 
average collapses onto a single well defined curve $G(x)$, see Eq.~(\ref{eq:no7}).
This is shown for system sizes $L = 64, 128, 256$ and $512$.   The number of systems in each ensemble is $10,000$.}
\label{F:H_t_random}
\end{figure}

A data collapse of the data using these values is shown in Fig.~\ref{F:H_t_random}.
It is in good agreement with Eq.~(\ref{eq:no7}), thus supporting our explanation for the appearance of a time dependence 
in $\langle H \rangle (t,L)$.
However, this is only a crude calculation as there are many effects we have ignored.
For instance, we have only calculated the average time for avalanches to reach the perturbed
site.
There is actually a distribution of times and this will contribute to the time dependence
of $\langle H \rangle (t,L)$.
In conclusion, we have analyzed damage spreading in the Oslo model, showing that damage is unable to evolve in a perturbed system.
The damage is statistically time invariant and scale free and thus allows for arbitrarily large values in infinite systems.
This phenomenon is due to the Abelian property of the Oslo model,
and this generalizes our result to all other models with Abelian properties, including the BTW model \cite{BTW87}, 
the Manna model \cite{Manna} and the model for interface depinning in a random medium \cite{MPacz, qEW}.
Thus, many of the classic models of SOC may be considered as lying on the edge of chaos \cite{PBak, Kauffman}.
Finally, we have shown how simulations may in fact lead to a time dependent ensemble
averaged damage and have calculated this for the case of random placement of the 
perturbative grain.

{\em Acknowledgments:}
We acknowledge very helpful discussions with Gunnar Pruessner, Nicholas Moloney and Henrik Jensen.
K. C. gratefully acknowledges the financial support of U.K. EPSRC through grant No. GR/R44683/01.

\end{document}